\documentstyle[12pt]{article}
\newcommand{\stk}[1]{\stackrel{*}{\overline}}

\begin{document}
\begin{center}
\vfill
\large\bf{Supersymmetry on the Surface $S_2$}
\end{center}
\vfill
\begin{center}
D.G.C. McKeon\\
Department of Applied Mathematics\\
University of Western Ontario\\
London\\
CANADA\\
N6A 5B7
\end{center}
\vfill
email: DGMCKEO2@.UWO.CA \\
Tel: (519)661-2111, ext. 88789\\
Fax: (519)661-3523
\eject
\section{Abstract}

By using symplectic Majorana spinors as Grassmann coordinates in a superspace
associated with the supersymmetric extension of the isometry group on the
spherical surface $S_2$, it proves possible to formulate supersymmetric
models on $S_2$ using superspace techniques.

\section{Introduction}

The formulation of gauge theories on a spherical surface [1] has provided insights into their
properties. The kinetic term for all fields on this surface involves the operator
$L_{\mu\nu} \equiv -x_\mu\partial_\nu + x_\nu \partial_\mu$, which is the generator
of the isometry group on the spherical surface.

In formulating a supersymmetric extension of the isometry group on any surface
of constant curvature, one must introduce a Fermionic operator $Q$ which is the ``square
root'' of the full angular momentum operator $J_{\mu\nu}$, whose commutation relation
is the same as that of $L_{\mu\nu}$. Since $J_{\mu\nu}$ does not commute
with $Q$ (in contrast to the translation operator $P_\mu$ in flat space
[2]), closure of the algebra often requires introduction of further
Bosonic
operators that act as internal symmetry generators (which do not commute
with $Q$) [3-5].

Supersymmetric models on several surfaces of constant curvature have
been formulated using $L_{\mu\nu}$ directly in the kinetic term for
component fields. In particular, supersymmetric models on $S_2$ and
$AdS_2$ have been devised [6]. (This approach is distinct from
supersymmetric models on a surface of constant curvature formulated by
specializing the gravitational background field in a supergravity model;
this approach has been used in [7].) The component field action for
a supersymmetric model on $S_2$ is [6]
$$S = \int \frac{dA}{a^2} \left\lbrace \left[\frac{1}{2} \psi^\dagger (x)
\left(\tau \cdot L + \zeta\right) \psi(x) - \phi^* \left( L^2 + \zeta (1
- \zeta)\right)\phi\right.\right.\nonumber$$
$$\left. - \frac{1}{2} F^*F\right] + \lambda_N \Big[2(1-
2\zeta)\phi^*\phi\nonumber$$
$$\left.\left. - (F^*\phi + F\phi^*) -
\psi^\dagger\psi\right]^N\right\rbrace .\eqno(1)$$
In (1), $\phi$ and $F$ are complex scalars and $\psi$ a two component
Dirac spinor, defined on the surface of a sphere of radius $a$ in three
dimensions. The angular momentum vector is $L = -i x \times \nabla$, the
$\tau$ are the Pauli spin matrices, and $\zeta$ and $\lambda_N$ are
arbitrary real parameters.

The action of (1) is invariant under the supersymmetry transformations
$$\delta \phi = \xi^\dagger \psi\eqno(2a)$$
$$\delta \psi = \left[2(\tau \cdot L + 1 - \zeta)\phi -
F\right]\xi\eqno(2b)$$
$$\delta F = -2 \xi^\dagger (\tau \cdot L + \zeta)\psi \eqno(2c)$$
and the special transformation
$$\delta \phi = \lambda i \left[2(1 - \zeta)\phi -
F\right]\eqno(3a)$$
$$\delta \psi = \lambda i \left[1 + 2\tau \cdot L\right]\psi\eqno(3b)$$
$$\delta F = \lambda i \left[-4 \left( L^2 + \zeta (1-\zeta)\right)\phi +
2 \zeta F\right]\eqno(3c)$$
as well as the usual rotations generated by the angular momentum
operator $\vec{J}$. These transformations are generated by
$\exp\left[\xi^\dagger Q - Q^\dagger \xi + i \lambda Z + i\vec{\omega}
\cdot \vec{J}\,\right]$ where
$Q$, $Q^\dagger$, $Z$ and $J^a$ satisfy the algebra
$$\left\lbrace Q_i, Q_j^\dagger\right\rbrace = Z\delta_{ij} - 2
\tau_{ij}^a J^a\eqno(4a)$$
$$\left[ J^a, Q_i^a\right] = -\frac{1}{2} \tau_{ij}^a Q_j\eqno(4b)$$
$$\left[ Z, Q_i\right] = -Q_i\eqno(4c)$$
$$\left[ J^a , J^b \right] = i\epsilon^{abc} J^c .\eqno(4d)$$
(For two other superalgebras associated with $S_2$, see ref. [5].)
This algebra has a representation in superspace
$$Q = (\tau \cdot x + \beta) \frac{\partial}{\partial\theta^\dagger}
 + \left( \frac{\partial}{\partial \beta } - \tau \cdot \nabla
\right)\theta \eqno(5a)$$
$$Q^\dagger = \frac{\partial}{\partial\theta}(\tau \cdot x + \beta) -
\theta^\dagger
\left( \frac{\partial}{\partial \beta } - \tau \cdot \nabla
\right) \eqno(5b)$$
$$J^a = \frac{1}{2}\left[\frac{\partial}{\partial\theta}
\tau^a \theta + \theta^\dagger \tau^a
\frac{\partial}{\partial \theta^\dagger }\right] + L^a \eqno(5c)$$
$$Z = \left(\theta^\dagger \frac{\partial}{\partial\theta^\dagger} -
\theta^T \frac{\partial}{\partial\theta^T}\right)\eqno(5d)$$
where $\beta$ is a Bosonic variable with no apparent physical
significance and $\theta$ is a Dirac spinor that acts as a Grassmann
coordinate.

A component field model similar to (1) has been formulated as a surface
in $2 + 1$ dimensions associated with the space $AdS_2$ [6,8]. For this
space (as well as $AdS_3$) it has also proved possible to formulate
supersymmetric models in superspace [8,9]. This has been feasable as the
Grassmann coordinates in $AdS_2$ and $AdS_3$ are spinors with two
independent components (Majorana spinors for $AdS_2$ and Majorana-Weyl
for $AdS_3$) which limits the component fields which can contribute to a
scalar superfield to being a pair of real scalars and a two component
Majorana spinor.  The superfield actions that are devised have viable
kinetic and interaction contributions for these component fields.
Curiously, the superfield actions on $AdS_2$ are distinct from the
component field action that resembles the $S_2$ actions of eq. (1).

Though in eq. (5) we have a representation of the supersymmetry
operators in superspace, it is not immediately clear how to construct a
superfield model in this superspace.  A general superfield takes the
form
$$\begin{array}{l}
\Phi 
\left( x, \theta, \theta^\dagger \right) = \phi (x) +
\psi^\dagger (x) \theta + \theta^\dagger \psi(x)\\
\;\;\;\;\;\;\;\;\;\;\;\;\;\;\;\;\;+ F(x) \theta^\dagger 
\theta + V^a(x)\theta^\dagger \tau^a\theta +
\left(\lambda^\dagger (x)\theta + \theta^\dagger\lambda
(x)\right)\theta^\dagger\theta + G(x)\left(\theta^\dagger
\theta\right)^2,\end{array}\eqno(6)$$
where $\phi$, $F$ and $G$ are scalars, $\psi$ and $\lambda$ are spinors
and $V^a$ is a vector. Reducing the number of independent components in
$\Phi$ as is done in $3 + 1$ dimensional space does not seem feasable,
as there does not appear to be an analogue of the operators $D$ that
permit one to define chiral superfields. (The possibility of having a
real gauge superfield has not been persued.)

In the next section it is shown that by replacing the Dirac spinor
$\theta$ with a pair of symplectic Majorana spinors, one can write down
a suitable superfield action involving just one of these two spinors.
The only problem is that the action is not Hermitian; this problem is
rectified by adding to this action its Hermitian conjugate which
necessarily involves the second of the two symplectic Majorana spinors.

\section{Superfield action on $S_2$.}

We first note that as $\tau^2 \tau^a \tau^2 = -\tau^{aT}$,
a suitable charge conjugation matrix is provided by $C = \tau^2$, and
the charge conjugate of a spinor $\psi$ is $\psi_C = C\psi^{\dagger T} = (\tilde{\psi})^\dagger$.
Since $(\psi_C)_C = -\psi$ one cannot have a Majorana spinor in $3 + 0$
dimensions; one can however have a pair of symplectic Majorana spinors
$$\psi_1 = \left( \psi + \psi_C\right)/\sqrt{2}\eqno(7a)$$
$$\psi_2 = \left( \psi - \psi_C\right)/\sqrt{2}\eqno(7b)$$
so that
$$\left( \tilde{\psi}_1 \psi_1\right)^\dagger =
-\left(\tilde{\psi}_2 \psi_2\right)\eqno(8a)$$
$$\left(\psi_1\right)_C = -\psi_2\eqno(8b)$$
$$\left(\psi_2\right)_C = +\psi_1\eqno(8c)$$
(viz. $(\psi_\alpha)_C = - \epsilon_{\alpha\beta}\psi_\beta$). Upon
decomposing the spinorial generator $Q$ of eq. (4) in this way, the
anticommutator of eq. (4a) becomes
$$\left\lbrace Q_1, \tilde{Q}_1 \right\rbrace = -2 \tau \cdot J = -
\left\lbrace Q_2, \tilde{Q}_2 \right\rbrace \eqno(9a)$$
$$\left\lbrace Q_1, \tilde{Q}_2 \right\rbrace = Z = 
\left\lbrace Q_2, \tilde{Q}_1 \right\rbrace \eqno(9b)$$
where $\tilde{Q}_\alpha = Q_\alpha^T C = (Q_\alpha)^\dagger_C$. Using
symplectic Majorana spinors $\theta_1$ and $\theta_2$, a superspace
representation of $Q_\alpha$, $\tilde{Q}_\alpha$, by eq. (5), is given
by
$$Q_1 = \left(-\tau \cdot x \frac{\partial}{\partial \tilde{\theta}_1} +
\beta \frac{\partial}{\partial \tilde{\theta}_2}
\right) + \left(\tau \cdot \nabla \theta_1 - \frac{\partial}{\partial
\beta} \theta_2\right)\eqno(10a)$$
$$Q_2 = \left(\tau \cdot x \frac{\partial}{\partial \tilde{\theta}_2} -
\beta \frac{\partial}{\partial \tilde{\theta}_1}
\right) + \left(\tau \cdot \nabla \theta_2 - \frac{\partial}{\partial
\beta} \theta_1\right)\eqno(10b)$$
$$\tilde{Q}_1 = -\frac{\partial}{\partial{\theta}_1} \tau \cdot x
- \frac{\partial}{\partial{\theta}_2}\beta
 - \tilde{\theta}_1 \tau \cdot \nabla - \tilde{\theta}_2
\frac{\partial}{\partial\beta}\eqno(10c)$$
$$\tilde{Q}_2 = \frac{\partial}{\partial{\theta}_2} \tau \cdot x
+ \frac{\partial}{\partial{\theta}_1}\beta
 - \tilde{\theta}_2 \tau \cdot \nabla - \tilde{\theta}_1
\frac{\partial}{\partial\beta}\eqno(10d)$$
$$Z = -\left(\theta_1^T 
\frac{\partial}{\partial{\theta}_2^T} + 
\theta_2^T \frac{\partial}{\partial\theta_1^T}\right).\eqno(10e)$$
Excising the parts of $Q_a$ in (10) dependent on $\beta$ or
$\frac{\partial}{\partial\beta}$, we define
$$q_1 = - \tau \cdot x \frac{\partial}{\partial\tilde{\theta}_1} + \tau
\cdot \nabla \theta_1 = + \left(q_2 \right)_C\eqno(11a)$$
$$q_2 = \tau \cdot x \frac{\partial}{\partial\tilde{\theta}_2} + \tau
\cdot \nabla \theta_2 = - \left(q_1 \right)_C\eqno(11b)$$
$$\tilde{q}_1 = -\frac{\partial}{\partial\theta_1} \tau
\cdot x - \tilde{\theta}_1 \tau \cdot \nabla \eqno(11c)$$
$$\tilde{q}_2 = \frac{\partial}{\partial\theta_2} \tau
\cdot x - \tilde{\theta}_2 \tau \cdot \nabla \eqno(11d)$$
so that
$$\left\lbrace q_1, \tilde{q}_1\right\rbrace = -2 \tau^a J_1^a
\eqno(12a)$$
$$\left\lbrace q_2, \tilde{q}_2\right\rbrace = +2 \tau^a J_2^a
\eqno(12b)$$
$$\left[ J_\alpha^a, J_\alpha^b \right] = i\epsilon^{abc} J_\alpha^c
\;\;\;\;(\alpha = 1, 2)\eqno(12c)$$
$$\left[ J_\alpha^a, q_\alpha \right] = -\frac{1}{2}
\tau^a q_\alpha \;\;\;\;\; (\alpha = 1, 2)\eqno(12d)$$
where
$$J_\alpha^a = -i(x \times \nabla)^a + \frac{1}{2}
\frac{\partial}{\partial \theta_\alpha}\;\tau^a \theta_\alpha\;
.\eqno(13)$$

We note that [8] with $Q$ given by (5a),
$$\left[Q, \Delta\right] = 0 = \left[ Q, R^2\right]\eqno(14)$$
where
$$\Delta = \theta^\dagger \,\frac{\partial}{\partial\theta^\dagger} + 
\theta \,\frac{\partial}{\partial\theta} + x \cdot \nabla + \beta\, 
\frac{\partial}{\partial\beta} \eqno(15a)$$
$$R^2 = x^2 - \beta^2 - 2\theta^\dagger \theta \; .\eqno(15b)$$
Similarly, we find that for $\alpha = 1, 2$
$$\left[ q_\alpha , \Delta_\alpha \right] = 0 = \left[ q_\alpha,
R_\alpha^2\right]\eqno(16)$$
where
$$\Delta_1 = \theta_1 \,\frac{\partial}{\partial\theta_1} + x \cdot
\nabla ,\;\;\;\;\; \Delta_2 = \theta_2 \,\frac{\partial}{\partial\theta_2} + x \cdot
\nabla\eqno(17a)$$
$$R_1^2 = x^2 + \tilde{\theta}_1\theta_1\;\;\;\;\;\;\;
R_2^2 = x^2 - \tilde{\theta}_2 \theta_2 \; .\eqno(17b)$$
$$ \;\;\;\;\;\;= \left(R_2^2\right)^\dagger\;\;\;\;\;\;\;\;\;\;\;=
\left(R_1^2\right)^\dagger \nonumber$$

We now introduce two superfields
$$\Phi_1\left(x,\theta_1\right) = \left(\Phi_2\left(x,
\theta_2\right)\right)^\dagger = \phi (x) + i\tilde{\psi}_1(x)\theta_1 +
i F(x) \tilde{\theta}_1\theta_1\eqno(18a)$$
$$\Phi_2\left(x,\theta_2\right) = \left(\Phi_1\left(x,
\theta_1\right)\right)^\dagger = \phi^* (x) + i\tilde{\psi}_2(x)\theta_2 +
i F^*(x) \tilde{\theta}_2\theta_2\eqno(18b)$$
where $\phi$ and $F$ are complex scalars and $\psi_1$ and $\psi_2$ are a pair of
symplectic Majorana spinors. In addition, we define the operators
$$e_1 (\alpha , \beta ) = -\alpha \tau \cdot x \,
\frac{\partial}{\partial\tilde{\theta}_1} + \beta \tau \cdot \nabla
\theta_1 = +(e_2)_C\eqno(19a)$$
$$e_2 (\alpha , \beta ) = \alpha \tau \cdot x \,
\frac{\partial}{\partial\tilde{\theta}_2} + \beta \tau \cdot \nabla
\theta_2 = -(e_1)_C\eqno(19b)$$
$$\tilde{e}_1 (\alpha , \beta ) = - \alpha
\,\frac{\partial}{\partial\theta_1} \,\tau \cdot x - \beta
\tilde{\theta}_1 \tau \cdot \nabla\eqno(19c)$$
$$\tilde{e}_2 (\alpha , \beta ) =  \alpha
\,\frac{\partial}{\partial\theta_2} \,\tau \cdot x - \beta
\tilde{\theta}_2 \tau \cdot \nabla ,\eqno(19d)$$
where $\alpha$ and $\beta$ are real constants.

Under a transformation generated by $q_1$, we find that
$$\delta \Phi_1 = \left[\tilde{\epsilon}_1 q_1 , \Phi_1\right] = 
\left[ -i\tilde{\epsilon}_1 \tau \cdot x \psi_1\right] + \left[ -2i F
\tilde{\epsilon}_1 + \tilde{\epsilon}_1 \tau \cdot \partial
\phi\right]\theta_1\nonumber$$
$$+ \left[ - \frac{i}{2} \tilde{\epsilon}_1 \tau \cdot \partial
\psi_1\right]\tilde{\theta}_1\theta_1\eqno(20)$$
from which we can deduce the changes in $\phi$, $\psi$ and $F$. The
change in the $\tilde{\theta}_1\theta_1$ contribution to $\Phi_1$ is a
total derivative, and consequently an action invariant under the
supersymmetry transformation of (20) is given by
$$S_1 = i \int d^3x \int d^2\theta_2 \delta^2\left(\theta_2\right) \int d^2\theta_1
\delta\left(R_1^2 - a^2\right) \Phi_1 \tilde{e}_1 e_1 \Phi_1 \;
.\eqno(21)$$
In (21) we first note that we have defined $\theta$-integration so that
$$\int d^2\theta_\alpha \tilde{\theta}_\alpha \theta_\alpha =
1\;\;\;\;\;\; (\alpha = 1, 2).\eqno(22)$$
The $\delta$-functions are taken to be
$$\delta^2 \left(\theta_1\right) = \tilde{\theta}_1 \theta_1 = -
\left(\tilde{\theta}_2 \theta_2\right)^\dagger = -
\left[\delta\left(\theta_2\right)\right]^\dagger\eqno(23)$$
$$\delta \left(R_1^2 - a^2\right) = \delta\left(x^2 - a^2\right) +
\tilde{\theta}_1 \theta_1 \delta^\prime \left(x^2 - a^2\right)\eqno(24)$$
$$\;\;\;\;\;\;\;\;\;\;= \delta\left(x^2 - a^2\right) \left[1 - 
\frac{1}{2a^2} \,\tilde{\theta}_1 \theta_1 \left(x \cdot \partial +
1\right)\right]\nonumber$$
$$= \left[ \delta \left(R_2^2 - a^2\right)\right]^\dagger\; .\nonumber$$
The product of all of the contributions to the integrand of eq. (21) is
necessarily of the form of the superfield of eq. (18a), and hence under
the transformation of eq. (20), the integrand transforms as a total
derivative. The action $S_1$ is consequently invariant under the
transformations of eq. (20). However, the action is not Hermitian. Our
full action is taken to be
$$S_{(0)} = S_1 + S_2\eqno(25)$$
where
$$S_2 = S_1^\dagger = -i\int d^3x \int d^2\theta_1
\delta^2\left(\theta_1\right)\int d^2\theta_2 \delta\left(R_2^2 -
a^2\right)
\Phi_2 \tilde{e}_2 e_2 \Phi_2 .\eqno(26)$$
It is evident that the action $S_2$ is invariant under transformations
generated by $q_2$. Furthermore, it is possible to supplement the action
$S_{(0)}$ of eq. (25) with interactions of the form
$$S_{(N)} = \lambda_N \int d^2\theta_1
d^2\theta_2\left[\delta^2\left(\theta_2\right)\delta\left(R_1^2 -
a^2\right)\Phi_1^N- \delta^2\left( \theta_1\right)\delta\left(R_2^2 -
a^2\right)\Phi_2^N\right]\;\;\;\;\;\;(N = 2, 3 \ldots)\eqno(27)$$
where $\lambda_N$ is a coupling.

It is now feasable to determine the component field form of the action.
This entails being able to define $\Phi_\alpha$ off the spherical
surface $S_2$. To do this, we employ the invariant conditions
$$\Delta_\alpha \Phi_\alpha = \omega \Phi_\alpha\;\;\;\;\;(\alpha = 1,
2)\eqno(28)$$
with $\Delta_\alpha$ defined in (17a) and $\omega$ being a real
constant. It is easy to establish that (28) implies that
$$x \cdot \partial \phi = \omega \phi\eqno(29a)$$
$$x \cdot \partial \psi_\alpha = (\omega - 1)\psi_\alpha \;\;\;\;(\alpha
= 1, 2)\eqno(29b)$$
$$x \cdot \partial F = (\omega - 2)F\; .\eqno(29c)$$
We also use $\partial^2 = \frac{1}{x^2} \left(-L^2 + (x \cdot \partial
)^2 + (x \cdot \partial )\right)$.
It is now possible to show that
$$S_{(0)} = \int d^3x \,\delta\left(x^2 - a^2\right) \Bigg\{ 2\left(-\alpha\beta + \alpha^2
\omega\right)\left(\phi F + \phi^*F^*\right) -
2i\alpha^2a^2\left(F^2\right)\nonumber$$
$$\left. -F^{*2}\right) + i\alpha\beta \left[\tilde{\psi}_1\left(\tau
\cdot L + \frac{3}{2}\right)\psi_2 + \tilde{\psi}_2\left(\tau \cdot L +
\frac{3}{2}\right)\psi_2\right]\eqno(30a)$$
$$\left. + \frac{i\beta^2}{a^2} \left[\phi\left(L^2 - \omega(\omega +
1)\right)\phi -\phi^* \left(L^2 - \omega(\omega + 1)\phi^*\right)\right] +
\frac{2i\alpha\beta\omega^2}{a^2}\left(\phi^2 -
\phi^{*2}\right)\right\rbrace\nonumber$$
$$\!\!\!\!\!\!\!\!\!\!\!\!S_{(2)} = \lambda_2 \int d^3x\, \delta\left(x^2 - a^2\right) 
\Big\{ 2i\left(\phi F - \phi^*F^*\right) \nonumber$$
$$\left. + \frac{1}{2} \left( - \tilde{\psi}_1 \psi_1 +\tilde{\psi}_2
\psi_2\right)
- \frac{1}{2a^2} (2\omega + 1) \left(\phi^2 +
\phi^{*2}\right)\right\rbrace\eqno(30b)$$
$$S_{(3)} = \lambda_3 \int d^3x \,\delta \left(x^2 - a^2\right) \left\lbrace
3i\left( F\phi^2 - F^*\phi^{*2}\right) - \frac{3}{2}
\left(\phi\tilde{\psi}_1\psi_1 -
\phi^*\tilde{\psi}_2\psi_2\right)\right.\nonumber$$
$$\left. - \frac{1}{2a^2} (3\omega + 1)\left(\phi^3 +
\phi^{*3}\right)\right\rbrace  \,.\eqno(30c)$$
The expressions for $S_{(N)}$ for $N > 3$ can easily be generated in the
same manner.

\section{Summary}
We have demonstrated in this paper how a superfield formalism can be
used to construct supersymmetric models on $S_2$ associated with algebra
of eq. (4). The resulting model, whose component field form is given in
eq. (30), is quite distinct from the component field model of eq. (1).

It is quite easy to formulate superfield models associated with the
superalgebra on $S_2$
$$\left\lbrace Q, Q^\dagger\right\rbrace = Z + 2\tau \cdot J\eqno(31a)$$
$$\left[ J^a, Q\right] = -\frac{1}{2} \tau^a Q\eqno(31b)$$
$$\left[Z, Q\right] = Q\eqno(31c)$$
$$\left[J^a, J^b\right] = i\epsilon^{abc}J^c\eqno(31d)$$
as it is so akin to the algebra of eq. (4). The superalgebra [5]
$$\left\lbrace Q, \tilde{Q}\right\rbrace = \tau \cdot J\;\;\;\;\;\;
\left\lbrace Q, Q^\dagger\right\rbrace = \tau \cdot Z
\eqno(32a)$$
$$\left[ J^a, Q\right] = -\frac{1}{2} \tau^a Q\;\;\;\;\;\;\;
\left[ Z^a, \tilde{Q}\right] = \frac{1}{2} Q^\dagger \tau^a\eqno(32b)$$
$$\left[ J^a, J^b\right] = i\epsilon^{abc}J^c\;\;\;\;\;\;
\left[ Z^a, Z^b\right] = -i\epsilon^{abc}J^c\eqno(32c)$$
on $S_2$ is quite distinct from those of eqs. (4) and (31); a model
invariant under transformations related to this algebra are more likely
to be difficult to devise. It would also be interesting to discover how
the superfield formalism could be used to compute radiative effects on
$S_2$. These matters are currently under consideration.

\section{Acknowledgements}
NSERC provided financial support. Conversations with C. Schubert and T.N.
Sherry were most helpful. R. and D. MacKenzie had useful advice.

\end{document}